\documentclass[twocolumn,showpacs,preprintnumbers,amssymb,pre,superscriptaddress]{revtex4}

\usepackage{graphicx}
\usepackage{dcolumn}
\usepackage{bm}
\usepackage{amstext}

\begin{document}



\title{A Macroscopic Scale Model of Bacterial Flagellar Bundling}

\author{MunJu Kim, James C. Bird,
Annemarie J. Van~Parys, Kenneth S. Breuer,  Thomas R. Powers\\
\textit{Division of Engineering, Box D, Brown University,
Providence, RI 02912}} 

\date{August 14, 2003}


\begin{abstract}

\textit{Escherichia coli} and other bacteria use rotating helical
filaments to swim. Each cell typically has about four filaments,
which bundle or disperse depending on the sense of motor rotation.
To study the bundling process, we 
built a macroscopic scale model consisting of stepper-motor-driven
polymer helices in a tank filled with a high-viscosity silicone
oil. The Reynolds number, the ratio of viscous to elastic
stresses, and the helix geometry of our experimental model
approximately match the corresponding quantities of the full scale
\textit{E. coli} cells. We analyze digital video images of the
rotating helices to show that the initial rate of bundling is
proportional to the motor frequency and is independent of the
characteristic relaxation time of the filament.  We also determine
which combinations of helix handedness and sense of motor rotation
lead to bundling.

\end{abstract}

\pacs{87.19.St, 87.16.Qp, 87.16.-b, 47.15.Gf}



\maketitle



\noindent\textbf{Introduction.} \textit{Escherichia coli} and
\textit{Salmonella typhimurium} cells have several helical
propellers, or flagella, which they use to swim. Each flagellum
consists of a rotary motor embedded in the cell wall, a short (50
nm) flexible hook which acts as a universal joint, and a helical
filament about 20 nm in diameter and about $10$ $\mu$m
long~\cite{MacNab1996}. The trajectory of an individual swimming
cell consists of runs interrupted by tumbles. For most of a run,
the motors turn counterclockwise when viewed from outside the
cell, the filaments wrap into a tight bundle, and the cell swims
along a roughly straight path.  Near the end of a run, one or more
of the motors reverses direction, the corresponding filaments come
out of the bundle, and the cell moves erratically, or tumbles. The
tumbling process is complex and involves polymorphic transitions
of the filament first from the left-handed ``normal" state to the
right-handed ``semi-coiled" state, and then to the ``curly-1"
state~\cite{turner_ryu_berg2000}. The first transition reorients
the cell body. When the motors resume their counterclockwise
rotation, the curly-1 filaments transform directly to the normal
state and rejoin the bundle, and the cell resumes its initial
speed~\cite{turner_ryu_berg2000}.

The chemotaxis strategy of {\it E. coli} is to decrease the
likelihood of tumbling during runs which happen to carry the cell
toward higher concentrations of chemoattractants.  Thus, the
formation and dispersal of the helical bundle is central to
bacterial chemotaxis. Since the radius of the flagellar filament
is well below optical wavelengths, and the motor rotation is
relatively rapid (100 Hz), it is difficult to study the mechanics
of the bundling process directly.  Therefore, we built a
macroscopic scale-model system consisting of flexible rotating
helices in a very viscous fluid. By including the viscous
fluid and properly accounting for the relative strengths of viscous
and elastic stresses, our scale-model builds upon
and extends the work of Macnab, who studied the
geometry of rotating flexible helices in a bundle~\cite{macnab1977}.

Our paper begins with a
discussion of the material parameters of bacterial flagella and
how we chose the parameters for the experimental model. The next
section describes the geometry of bundled helices and the symmetry
requirements for bundling, including helix handedness,
motor-rotation sense and relative speed, and phase relations.
Finally, we describe measurements of the characteristic time scale
governing the initial stages of the bundling process.

\begin{figure*}
\includegraphics[height=4in]{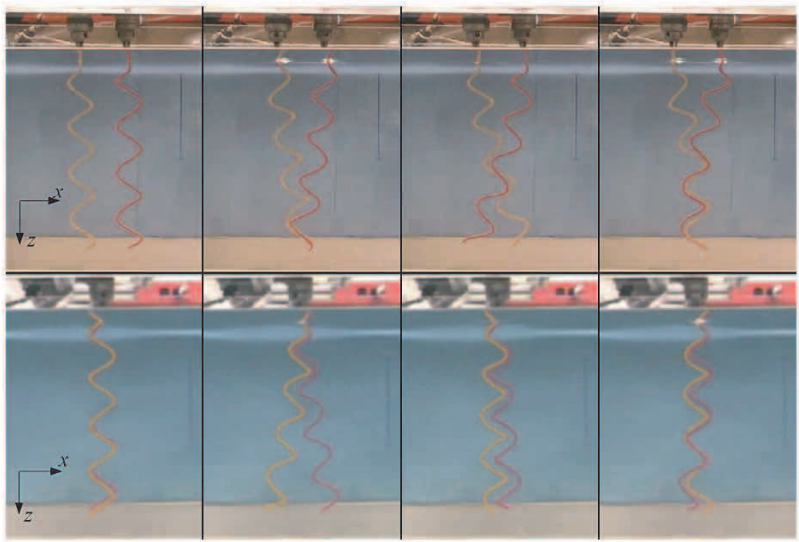}
\caption{Bundling sequence. The top panels show our left-handed
model ``semi-coiled" helices (see Table~\ref{polytable}) at $t=0$,
$96$, $168$, and $264$ s; the bottom panels show the helices at
the same times, viewed from the side. The scale bars are $100$ mm
long; the helices are $310$ mm long (from chuck to tip), $4.0$ mm
in diameter, and turning at $0.1$ Hz.} \label{bundlepicture}
\end{figure*}

\noindent\textbf{Scale Model.} \textit{E. coli} usually has
several filaments per cell, but for the purpose of studying
bundling it is simplest to consider the case with two filaments.
Unless the two helices are identical and placed on diametrically
opposite sides of the body (an unlikely occurrence), the
hydrodynamic torque on the rotating helices causes the cell body
to counter-rotate to make the total torque on the cell vanish.
This counter-rotation plays some role in bundle
formation~\cite{anderson1975,powers2002}, since it tends to wrap
the filaments around each other. However, this mechanism for
bundling does not lead to tight bundles~\cite{powers2002}.
Furthermore, bundles readily form from filaments of stationary
cells bonded to a surface~\cite{howardlinda_unpublished}.
Therefore, we disregard the effects of body rotation and
translation and consider two helices rotated by stationary motors.
We also neglect the effect of the cell body itself on the flow.

At the characteristic length and time scales of bacteria, viscous
effects dominate inertial effects~\cite{purcell1977}. To see why,
recall that the Reynolds number $Re=\rho v \ell/\eta$ determines
the relative importance of inertia to viscous stresses, where
$\rho$ is the density of the fluid, $\eta$ is the viscosity, $v$
is a characteristic flow velocity, and $\ell$ is a characteristic
length over which the flow varies. Using $30$ $\mu$m/s as a
typical swimming speed, $1$ $\mu$m as a typical cell-body size,
and the density and viscosity of water ($\rho\approx1000$
kg$/$m$^3$, $\eta\approx10^{-3}$ N-s/m$^2$) yields a Reynolds
number of $Re\approx3\times10^{-5}$. (Using the typical helix
radius $\ell\approx0.2$ $\mu$m and the linear azimuthal helix
velocity $v\approx 2\pi\times100$ Hz$\times\ell$ yields
approximately the same Reynolds number.)
At low Reynolds number, the flow field induced by a point force
falls off inversely with distance~\cite{happel_brenner1965},
leading to a long-range hydrodynamic interaction between two
rotating helices. (Since the net force on a \textit{swimming}
bacterium vanishes, the far field flow induced by a swimming
bacterium falls off faster than $1/r$, roughly like a dipole,
$1/r^2$.) Another characteristic of low Reynolds number flow is
that the drag per unit length on a long slender body depends
weakly on the filament diameter and the shape of the cross
section~\cite{lighthill1975}.
\begin{table}

\begin{tabular}{|l||c|c|l|}\hline
Form & Pitch, $\mu$m & Diameter, $\mu$m & handedness\\
\hline\hline
normal & 2.3 & 0.4 & left\\
semi-coiled & 1.1 & 0.5 & right\\
curly 1 & 1.0 & 0.3 & right\\
model normal&$11\times10^4$ & $2.54\times10^4$ & left\\
model ``semi-coiled"& $6.4\times10^4$ & $2.54\times10^4$ & right and left\\
\hline
\end{tabular}
\caption{Helix parameters: approximate pitch, approximate
diameter, and handedness for some polymorphic forms of flagellar
filaments from~\cite{turner_ryu_berg2000}
and~\cite{howardlinda_unpublished}. We also include the dimensions
of our model helices. For the models, the diameter is the diameter
of the mandrel.} \label{polytable}
\end{table}

Elastic stresses balance the viscous stresses on rotating
bacterial flagella, causing the filaments to bend and twist as
they wrap into a bundle.  Filaments can also undergo polymorphic
transformations when the motors turn clockwise; for simplicity, we
do not attempt to include polymorphic transformations in our
macroscopic model. The bending modulus $A=EI$ determines the
resistance of an elastic rod to bending, where $E$ is the Young's
modulus and $I$ is the moment of inertia of the cross
section~\cite{landau_lifshitz_elas}.  There are few measurements
of the bending resistance, with reported values ranging from
$10^{-24}$ N-m$^2$~\cite{fushime_maruyama_asakura1972} to
$10^{-22}$ N-m$^2$~\cite{hoshikawa_kamiya1985}. Likewise, the
value of twist modulus $C$ of the filament is not precisely
characterized, although the twist compliance of approximately
$0.2$ $\mu$m-long filaments attached to the hook and a locked
motor has been measured~\cite{blockblairberg1989}. Since it is
difficult to separate the twist modulus of the hook from the
modulus of the filament in this measurement, we will simply assume
that $C\approx A$ for the filament. For a filament with axial
length $L$, the product of the characteristic filament relaxation
time and the motor angular velocity, $M=\eta\omega L^4/A$,
determines the importance of viscous drag relative to flexibility
(see~\cite{machin1958} and~\cite{wiggins_goldstein1998}). Since
$A=EI\propto Ea^4$, where $a$ is the filament radius, $M$ depends
sensitively on the aspect ratio $L/a$. To estimate $M$ for the
bacteria, we will use $L=7$ $\mu$m, the typical length of the
filaments of reference~\cite{turner_ryu_berg2000}. Using
$\omega=100\times2\pi$ rad/s, $\eta=10^{-3}$ N-s/m$^{2}$, and
$A=10^{-23}$ N-m$^2$ yields $M\approx150$ for the bacterial
filaments.
Note 
that the persistence length~\cite{doi_edwards1986}
$\xi_\mathrm{P}=A/(k_\mathrm{B}T)\approx 2.5$ mm is large,
allowing us to disregard thermal fluctuations.

There are three dimensionless numbers which determine the helix
shape: $P/L$, $R/L$, and $a/L$, where $P$ is the helical pitch, and $R$
is the helical radius. Note that the ratio $R/P$
determines the pitch angle $\alpha$ \textit{via} $\tan\alpha=2\pi
R/P$, and that $P$ and $L$ determine the number of turns (which
need not be an integer). If we consider two helices with initially
parallel axes, then the final geometrical parameter of interest is
the spacing $h$ between the two motors.

\noindent\textbf{Methods.} Figure~\ref{bundlepicture} shows the
macroscopic model. Two stationary stepper motors under independent
control rotate flexible polymer helices in a tank of silicone oil.
The viscosity of the silicone oil is $100$ N-s/m$^2$, or about
$10^5$ times the viscosity of water. The motors typically rotate
at about $0.1$ Hz, and the helix radius is about $13$ mm. For
these parameters, the Reynolds number $Re\approx10^{-3}$, not
quite as low as $10^{-5}$ but low enough to justify our neglect of
inertial effects. To minimize the effects of the walls of the tank
on the flow near the helices, all experiments are carried out with
the helices near the center of the tank. The base of the tank is
420 mm by 420 mm, and the depth of the silicone oil is 330 mm. The helices are made by
wrapping a Tygon$^{\mathrm{TM}}$ tube around a cylindrical mandrel
and filling the tube with epoxy. Food coloring is added to provide
contrast in the video images. Once the epoxy hardens, the tube
maintains its helical shape but is also flexible. We vary the
flexibility by varying the radius of the Tygon$^{\mathrm{TM}}$
tube. To get appropriate values of $M$ in our model, we use
helices with a filament radius of a few millimeters. The bending
stiffness of each helix is approximately determined by measuring
the deflection under a known force of a \textit{straight} tube
filled with hardened epoxy; for $2a=4.0$ mm, $4.7$ mm, and $5.6$
mm, we found $A=(3.46\pm0.14)\times10^{-3}$ N-m$^2$,
$(10.53\pm0.14)\times10^{-3}$ N-m$^2$, and
$(22.50\pm0.14)\times10^{-3}$ N-m$^2$, respectively. As in the case of the bacterial
filaments, we assume $C\approx A$.
Since the aspect ratio $L/a$ for our model is much smaller than that of
the full-scale bacterial flagella, our model is not a true scale model.
However, since the drag of a
slender body depends only weakly on the filament
radius~\cite{lighthill1975}, it is not necessary to attain exact
geometric similarity. We used two representative helical
geometries for our filaments (see Table~\ref{polytable} for a
comparison of the dimensions of our model helices and bacterial
flagella).  We systematically varied the stiffness of the
``semi-coiled" model. Note that the (inner) radii of our model
helices are all the same, since both types of helices were formed
from the same mandrel, whereas the helical radius of the bacterial
filament varies from polymorph to polymorph. All experiments
reported here were carried out with a motor shaft-to-shaft
separation of $h=61$ mm, which would roughly correspond to the typical
spacing of one micron in the full scale bacteria.
To allay concerns about effects of fatigue and
plastic deformation of the helices during the experiments, we
compared the deflection of the helices under a known weight before and
after some representative trials and found that these effects were
insignificant.

\begin{figure}
\includegraphics[height=1.5in]{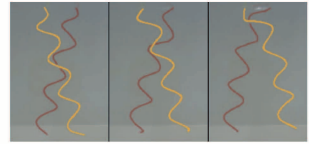}
\caption{(a): Sequence of left-handed helices turning clockwise at
$0.075$ Hz, at times $t=102$ s, $t=119$ s, and $t=146$ s. The
helices cross, and the crossing point migrates up to the motor,
where the helices periodically distort to escape a jam.}
\label{bundle_no_bundle}
\end{figure}
\noindent\textbf{Bundling in the Scale Model.}
Figure~\ref{bundlepicture} shows a sequence of snapshots of two
rotating left-handed helices (with four turns each) at
$M\approx170$ (see Movie 1, which is published as supporting
information on the PNAS website, {\tt www.pnas.org}). The induced
flows cause large deflections, and a bundle forms when the motors
turn counterclockwise. Note that the helices wrap around each
other in a right-handed sense; the flow field generated by each
helix tilts the other helix, causing the helices to roll around
each other and form a right-handed wrapping. In contrast with the
case of two helices, the deflection of a single rotating helix
with $M\approx170$ is small, with the axial compression or
extension less than ten percent of the axial length.  But an
understanding of the flow induced by a single helix sheds light on
the flows that cause bundling. Consider a left-handed helix. As
the helix rotates counterclockwise as viewed from the helix-side
of the motor, apparent helical waves travel along the filament
away from the motor and push fluid elements in the same direction.
These fluid elements also rotate about the vertical axis of the
helix in the same sense as the helix, counterclockwise, leading to
a right-handed trajectory for each fluid element.

Early scale-model experiments by Macnab using steel helices in air
showed that left-handed helices twisted around each other in a
right-handed manner can rotate indefinitely without jamming, but
that left-handed wrapping leads to jamming for clockwise rotation
and unwinding for counterclockwise rotation~\cite{macnab1977}.
Macnab also pointed out that the state of lowest elastic stress of
two left-handed helices with right-handed intertwisting is one in
which the helices are coaxial, in phase, and in contact. Since
this interwound state is unstable in the absence of external
torque, Macnab used a guide at the end of the bundle to keep the
helices from unravelling. In our experiment, viscous drag provides
the stabilizing torques that keep the helices interwound. Note
that the helices rotate against each other, leading to a region of
high shear between the filaments. When the motors turn
counterclockwise, we find that the bundle persists indefinitely as
long as the motor speeds are sufficiently low. If the motor speed
is greater than $0.1$ Hz, then the Y-shaped junction in the top
right panel of Fig.~\ref{bundlepicture} migrates up to the motors,
causing a jam.  Jams can also form when the motors turning the
helices in a steady-state bundle reverse; if a jam fails to form,
the filaments simply unwind.
\begin{figure}
\includegraphics[height=3in]{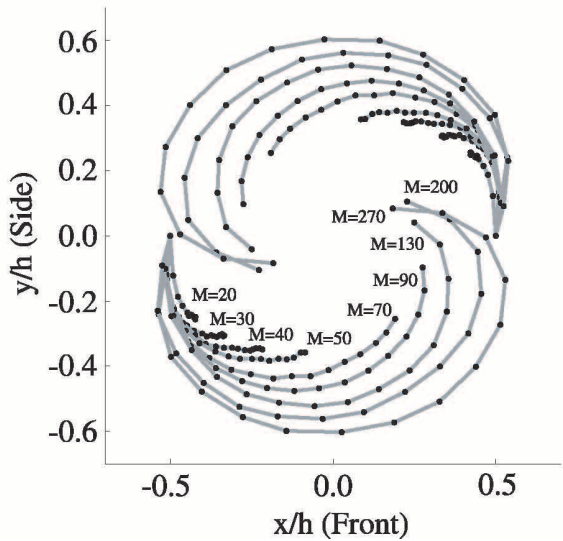}
\caption{ Trajectories of (projected) helix tips in the
$xy$-plane.  The spacing of the dots on each curve corresponds to
one revolution.} \label{spiral}
\end{figure}

In the example shown in Fig.~\ref{bundlepicture}, the two helices
have no initial phase difference. Since the motors run at a common
constant velocity and do not slip, the phase difference between
the motor shafts is constant. If there is an initial phase
difference, then the helices must bend and twist to attain the
in-phase interwound state. Typically one helix bends more than the
other, leading to a bundle which is tilted relative to the axis of
rotation of each motor. The motor characteristic in the scale
model, constant speed, differs from that of the bacterial motor,
which for typical loads and temperatures is roughly constant
torque~\cite{berg_turner1993,washizu_et_al1993}.  Thus, the
phase-locking observed in the bacterial flagella may arise from a
different mechanism than that observed here.

In the scale model, it is crucial for the rotation speeds of the
two motors to be close for bundling to occur. If the motors speeds
differ by as little as $8\%$, left-handed helices rotated
counter-clockwise do not bundle. The helices cross each other,
with the faster helix bending more than the slower helix. The axis
of the faster helix waves back and forth, but never wraps around
the slower helix.


Since left-handed helices turning counter-clockwise bundle,
symmetry implies that right-handed helices turning clockwise will
also bundle. Indeed, bundles of semi-coiled or curly flagella have
been observed in fluorescently labeled
filaments~\cite{turner_ryu_berg2000}. On the other hand,
left-handed helices turning clockwise do not bundle. When the
motors driving a bundle reverse, the bundle unwinds. Once the
bundle has unwound, or if the (left-handed) helices are initially
stationary, then the flow induced by the clockwise rotation causes
the helices to roll around each other and wrap in a {\it
left-handed} sense. However, the wrapping is not nearly as tight
as the counter-clockwise sense. The helices cross, and the
crossing point migrates up to the motors; see
Fig.~\ref{bundle_no_bundle}. When the crossing point reaches the
motors, instead of jamming, the helices distort at the proximal
ends to allow the rotation to continue.
None
of the other combinations of handedness and sense of rotation led
to bundling in our scale-model experiment. Helices of the same
handedness but turning in opposite directions tilt away from each
other and also tilt away from the plane containing the initial
helix axes. Helices of opposite handedness turning in the same
sense cross and do not bundle; helices of opposite handedness
turning in opposite senses tilt away from each other.

\begin{table}

\begin{tabular}{|l||c|c|c|c|c|}\hline
Curve label & $\omega_1$, rad/s& $\omega_2$, rad/s & $\omega_3$, rad/s &$\omega_1/\omega_2$  & $\omega_2/\omega_3$\\
\hline\hline
I & 0.314 & 0.707 & & 0.44 & \\
II & 0.393 & 0.943 & & 0.42 & \\
III &  & 1.17 & 2.12 & & 0.56\\
IV & 0.707 & 1.65 & 2.83 & 0.43 & 0.41 \\
V & 1.18 & 2.36 & & 0.50 & \\
VI & 1.65& & & &  \\
VII & 3.06 & & & & \\
 \hline
\end{tabular}
\caption{Rotation speeds and ratios for the curves (labeled
I--VII) shown in Fig.~\ref{deflect_meas}. The columns labeled
$\omega_1$, $\omega_2$, and $\omega_3$ correspond to the $4.0$ mm,
$4.7$ mm, and $5.6$ mm tubes respectively. The empty spaces arise
because most of the overlapping sets of curves (except for case
IV) consist of just two curves.} \label{datatable}
\end{table}

\begin{figure}
\includegraphics[height=2.8in]{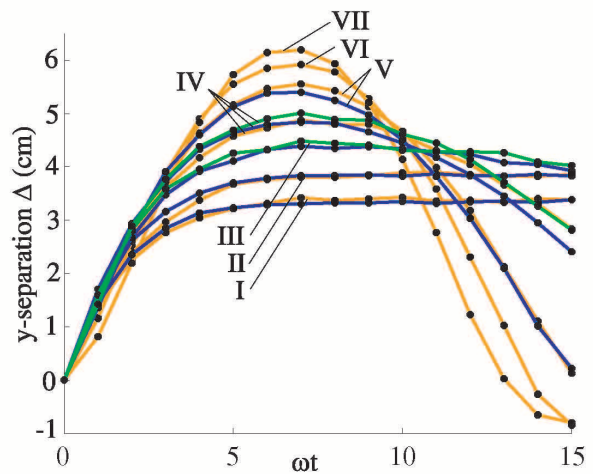}
\caption{Side view of (projected) tip-separation as a function of
time for rotation speeds and bending moduli; see
Table~\ref{datatable} for values.
On each curve, there is a black dot for every revolution. The gold
curves are for the 4.0 mm tube, the blue are for the 4.7 mm tube
and the green are for the 5.6 mm tube.} \label{deflect_meas}
\end{figure}
\noindent\textbf{Characteristic Time Scale.} There are two natural
time scales governing the dynamics of our system: the motor period
$2\pi/\omega$, and the elastic relaxation time scale $\eta L^4/A$.
To determine which combination of these controls the initial rate
of bundling, we measured the deflection of each helix as a
function of time. More precisely, we measured the position of the
projection of the tip of each helix onto the local axis of
rotation for the corresponding helix. Figure~\ref{spiral} shows
the trajectories of these points projected to the $xy$-plane. In
contrast to the case of Fig.~\ref{bundlepicture}, we used helices
with \textit{three} turns ($L=240$ mm) since with short helices it
is easier to attain small values of $M$. For $M<70$, the viscous
stresses induced by the rotating helices are too weak to deflect
the helices much. The helices cross when $M\approx70$, and begin
to wrap around each other when $M\approx130$. Since we focus on
the early stages of bundling, we do not show the complete
trajectories for moderate to large $M$ ($M>50$).
Defining $\Delta$ as the distance between ends of the helices in
the side view ($yz$-plane, where $z$ is parallel to the initial
helix axes, and $x$ is parallel to the line connecting the two
motors; Fig.~\ref{bundlepicture}), dimensional analysis implies
\begin{equation}
{\Delta\over h}=f(\omega t, M, Re, a/L, R/L, P/L, h/L).
\end{equation}
Since we approximate $C=A$, we omit $A/C$ from the list of
dimensionless groups. In this paper we do not report how $f$
depends on the geometrical parameters $R/L$, $P/L$, and $h/L$; we
present results for semi-coiled helices ($P/R\approx5$) with three
turns, $h=61$ mm, and $L=240$ mm.  We varied rotation speed
$\omega$ and filament radius $a$.  The function $f$ depends on
$\omega$ implicitly through $M$ and $Re$, and explicitly through
$\omega t$. Since $Re\ll1$, we can approximate $Re\approx0$ and
consider $f$ to depend on $\omega$ only through $M$ and $\omega
t$. Likewise, $f$ depends on $a$ implicitly through $M$ and
explicitly through $a/L$. The explicit $a/L$-dependence of $f$
arises from viscous drag because the Young's modulus $E=\pi a^4/4$
enters through $M$. Since the drag on a slender body depends
weakly on $a/L$, and since $M$ depends sensitively on $a/L$, we
consider $f$ to depend on $a$ only through $M$. Thus, for the
fixed geometrical parameters described above, we measure $f$ as a
function of $\omega t$ and $M$.

Figure~\ref{deflect_meas} shows $\Delta$ as a function of $\omega
t$ for various rotation speeds and bending moduli, and
Table~\ref{datatable} shows the speed corresponding to each
$\Delta$ \textit{vs.} $\omega t$ curve.
The ratio of the
speeds for the overlapping gold and blue curves is roughly
constant (see Table~\ref{datatable}). Note that the gold and blue
curves labeled V do not
overlap as closely as those labeled I, II, and III; correspondingly,
$\omega_1/\omega_2$ for the pair V noticeably differs from the
ratios for I, II, and III.  Likewise, the blue and green curves
labeled IV do not overlap as closely as those labeled III, leading to
the discrepancy between the $\omega_2/\omega_3$ values for pairs
III and IV.
Although $M$ systematically increases as the curve label varies
from I to VII, there is a significant spread in the values of $M$
for overlapping curves. This spread implies that our
identification of the bending modulus of a straight tube with that
of a helix with the same diameter tube leads to an uncertainty
greater than that of our measurements of $A$. Such an uncertainty
is not unexpected since we neglect the effects of the helix twist
modulus $C$ for simplicity. Despite the uncertainty in the helix
bending modulus, the data of Fig.~\ref{deflect_meas} and
Table~\ref{datatable} are strong evidence for the scaling behavior
of the bundling helices: two helices of different stiffness but
the same shape will yield the same $\Delta$ \textit{vs.} $\omega
t$ curves for the proper ratio of rotation speeds. In fact, if we
assume that the closely overlapping curves of
Fig.~\ref{deflect_meas} have the same $M$, then we can get a
better estimate of elasticity ratios using
$A_1/A_2=\omega_1/\omega_2$, since we can measure the rotation
speeds to high accuracy.

A characteristic time
of the early stages of the bundling process is the time at which
the separation $\Delta$ reaches a maximum.
Figure~\ref{deflect_meas} reveals that the position of this
maximum is proportional to the motor period, and depends only
weakly on the elastic relaxation time scale. Thus, in the early
stages of bundling, the helices follow the induced flow and the
motor period determines the characteristic time scale. The helices
begin to bundle after about six revolutions, in rough agreement
with the corresponding time for bacterial
flagella~\cite{turner_ryu_berg2000}.



\noindent\textbf{Conclusion.} Our macroscopic scale model
demonstrates that the bundling of bacterial flagella is a purely
mechanical phenomenon, arising from the interplay of hydrodynamic
interactions, bending and twisting elasticity, and geometry.
Our model allows us to study how the bundling phenomenon is
affected by parameters which are difficult to control in the
full-scale bacteria, such as the rate and direction of motor
rotation. For fixed values of the geometrical parameters, we
quantitatively characterized the early stages of the bundling
process, and found that the initial rate of bundling is mainly
determined by the motor period. Future work should address the role
of twisting vs. bending stiffness, the effect on bundling of the
translational and rotational flow fields induced by a swimming
bacterium, and bundle formation when there are more than two filaments.

\begin{acknowledgments}
We are grateful to H. Berg, N. Darnton, G. Huber, and  L. Turner
for discussions, and C. Bull, E. Chan,  F. Choi, S. Koehler, M.
Muller, and J. Simon for assistance and advice on constructing the
apparatus and collecting the data. This work is supported by NSF
grant no. CMS-0093658 (TRP) and the DARPA BioMolecular Motors Program
(KSB and TRP).
\end{acknowledgments}


\end{document}